\documentclass[twoside,a4paper,11pt]{proceedings}
\usepackage{graphicx}
\usepackage{hyperref}
\usepackage{movie15}
\usepackage{natbib}
\begin{document}
\pagenumbering{arabic}
\pagestyle{myheadings}
\thispagestyle{empty}
\vspace*{-1cm}
%{\flushleft\includegraphics[width=\textwidth,bb=58 650 590 680]{stamp.pdf}}
%{\flushleft\includegraphics[width=\textwidth,viewport=58 650 590 680]{stamp.pdf}}
{\flushleft\includegraphics[width=3cm,viewport=0 -30 200 -20]{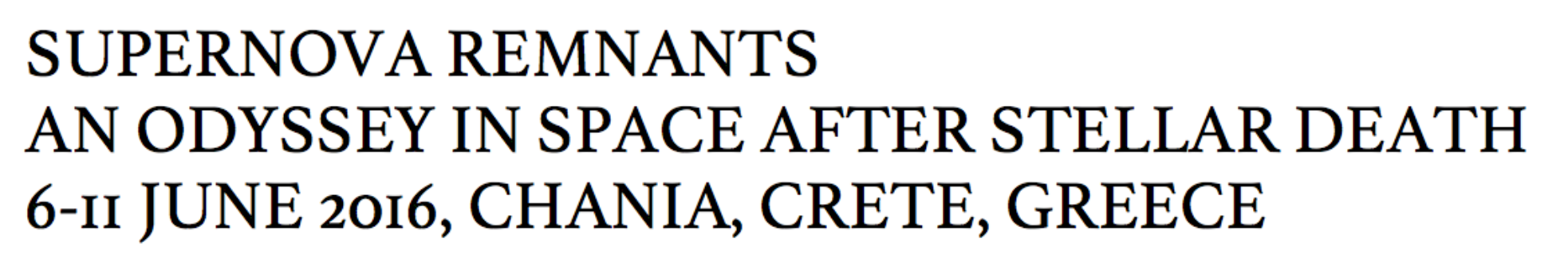}}
\vspace*{0.2cm}
\begin{flushleft}
{\bf {\LARGE
%%% TITLE of the paper. 
Supernova Remnant Evolution: from explosion to dissipation
}\\
\vspace*{1cm}
%%% Include here the LIST OF AUTHORS.
%%% Note that the last author has to be preceeded by an AND.
D.A. Leahy$^1$
%,
%
% Do not delete next few lines
}\\
\vspace*{0.5cm}
%
%%% AFFILIATIONS LIST.
%%% and the AFFILIATIONS LIST. Note that one affiliation per line.
%%% Add as many affiliations as necessary. 
$^{1}$
University of Calgary, Calgary, Alberta, Canada \\
%
% Do not delete next few lines
\end{flushleft}
% Headings
\markboth{
%%% Type the SHORT version of the paper t
Supernova Remnant Evolution
}{
%%%  First Author \& Second Author   OR   First-author et al. 
%%%  First Author \& Second Author   OR   First-author et al. if the author list contains three or more authors.
Leahy 
}
\thispagestyle{empty}
\vspace*{0.4cm}
\begin{minipage}[l]{0.09\textwidth}
\ 
\end{minipage}
\begin{minipage}[r]{0.9\textwidth}
\vspace{1cm}
\section*{Abstract}{\small
%%% Type the ABSTRACT of your paper
Here is considered the full evolution of a spherical supernova remnant. 
We start by calculating the early time ejecta-dominated stage and 
continue through the different phases of interaction with the circumstellar 
medium, and end with the dissipation and merger phase. 
The physical connection between the phases reveals new results. 
One is that the blast wave radius during the adiabatic phase is significantly
smaller than it would be, if one does not account for the blast wave
interaction with the ejecta.

\vspace{10mm}
\normalsize}
\end{minipage}
%%% BODY of the paper

\section{Introduction}

$\,\!$\indent A supernova remnant (SNR), the aftermath of a supernova explosion, is an important phenomenon of study in astrophysics. 
The typical $10^{51}$ erg of energy released in the explosion is transferred
primarily into the interstellar medium during the course of evolution
of a SNR. SNR are also valuable as tools to study the evolution of star, 
the evolution of the Galaxy, and the evolution of the interstellar medium. 
A SNR emits in X-rays from its hot shocked gas, in infrared from heated dust, 
and in radio continuum. 
The latter is via synchrotron emission from relativistic electrons accelerated at the SNR shock. 

The evolution of a single SNR can be studied and calculated using a hydrodynamics code. 
However to study the physical conditions of large numbers of SNR, it is
desirable to have analytic methods to obtain input parameters needed to run
a detailed hydrodynamic simulation.
The short paper describes the basic ideas behind the analytic methods, the 
creation of software to carry out the calculations and some new results of the 
calculations.

\section{Theory and calculation methods}

$\,\!$\indent The general time sequence of events that occur after a supernova
explosion, which comprise the supernova remnant can be
divided into a number of phases of evolution (Chevalier, 1977).
These are summarized as follows.

The ejecta dominated (ED) phase is the earliest phase when the ejecta from the explosion are not yet strongly decelerated by interaction. 
Self-similar solutions were found for the ejecta phase for the case of 
a supernova with ejecta with a power-law density profile
occurring in a circumstellar medium with a power-law density profile 
(Chevalier, 1982). 
Solutions were given for ejecta power-law indices of 7 and 12, and 
circumstellar medium  power-law indices of 0 and 2. 
The latter correspond to uniform a circumstellar medium and one caused
by a stellar wind with constant mass-loss rate.

The non-self similar evolution between ED to the Sedov-Taylor (ST) self-similar
phase was treated by Truelove and McKee (1999).
They found the so-called unified solution for the evolution of the forward
and reverse shock waves during this phase. 

The Sedov-Taylor (ST) self-similar phase is that for which the shocked ISM
mass dominates over the shocked ejecta mass and for which radiative energy
losses from the hot interior supernova remnant gas remain negligible.
These solutions are reviewed in numerous works, and are based on the original
work on blast waves initiated by instantaneous point energy injection in a 
uniform medium (Taylor, 1946; Sedov, 1946).

The next stage occurs when radiative losses from the post-shock gas become
important enough to affect the post-shock pressure and the dynamics of
expansion of the supernova remnant. 
This phase is called the pressure-driven snowplow phase (PDS phase).
Cooling sets in most rapidly for the interior
gas closest to the outer shock front, so that a thin cold shell forms behind the
shock. Interior to the thin shell, the interior remains hot and has significant
pressure, so it continues to expand the shell. 
The shell decelerates because it is gaining mass continually while 
being acted upon by the interior pressure. Here we refer the review of this
phase of evolution by Cioffi, McKee and Bertschinger (1988) 
This work also compares the analytic solutions to numerical hydrodynamic solutions for verification. 

When the interior pressure has dropped enough, it no longer influences the evolution of the massive cool shell. 
After this time, the supernova remnant is in the momentum conserving shell
(MCS phase.
The shell slows down according to the increase in swept up mass from the interstellar medium.
The final fate of a supernova remnant is merger with the interstellar medium,
when the shock velocity drops low enough the the expanding shell is no longer
distinguishable from random motions in the interstellar medium. 

To create an analytic model, or its realization in software, the different 
phases of evolution were joined. 
This problem is not simple, as pointed out in the work
of Truelove and McKee (1999). The evolution of the SNR is determined by
the distribution of mass, pressure and velocity within the SNR and the
shock jump conditions where there are any shocks. 
We follow similar methods to those in  Truelove and McKee (1999), to ensure
that the SNR evolution has continuous shock velocity and radius with time
and closely follows that of more detailed hydrodynamic calculations.

\section{Results}

$\,\!$\indent  Analytic solutions have been created 
which cover the evolution of the SNR from
early ED phase through ED-ST transition, ST phase, ST to PDS transition 
and final dissolution of the SNR.
We have taken care to properly join the different phases as noted above.
These solutions allow variation in the input physical parameters, 
such as explosion energy, ejected mass, ejecta and 
circumstellar medium density profiles and age. 
The numerical implementation of the solutions provides various output quantities, 
such as forward and reverse shock radius, and shock velocities and temperatures.
These can be compared to the observed properties of a given SNR.
Adjustment of the input parameters to match the observed properties 
yields estimates of the physical properties of the SNR, and also
allows estimates in uncertainties in these properties.

One of the new results from the analytic calculations is
that the shock radius at any given time during the ST phase is significantly
less than it is for the standard analytic ST solution. 
The reduced shock radius is a real physical effect and is understood 
as caused by interaction of the reverse shock wave with the 
(initially unshocked) ejecta. This result has not
been pointed out previously, and will change SNR parameter estimates
that have been made with the standard ST solution.  

Results of some of the calculations with the full-evolution model are
shown in Figures 1 and 2. 
Figure 1 shows the forward and reverse 
shock radii and velocities for the ED phase, ED to ST phase and ST phase,
for a SNR in a uniform circumstellar medium, and the parameters listed in the figure caption. 
Figure 2 shows similar plots for a SNR in a stellar wind circumstellar medium.

\begin{figure}
\center
\includegraphics[width=\textwidth]{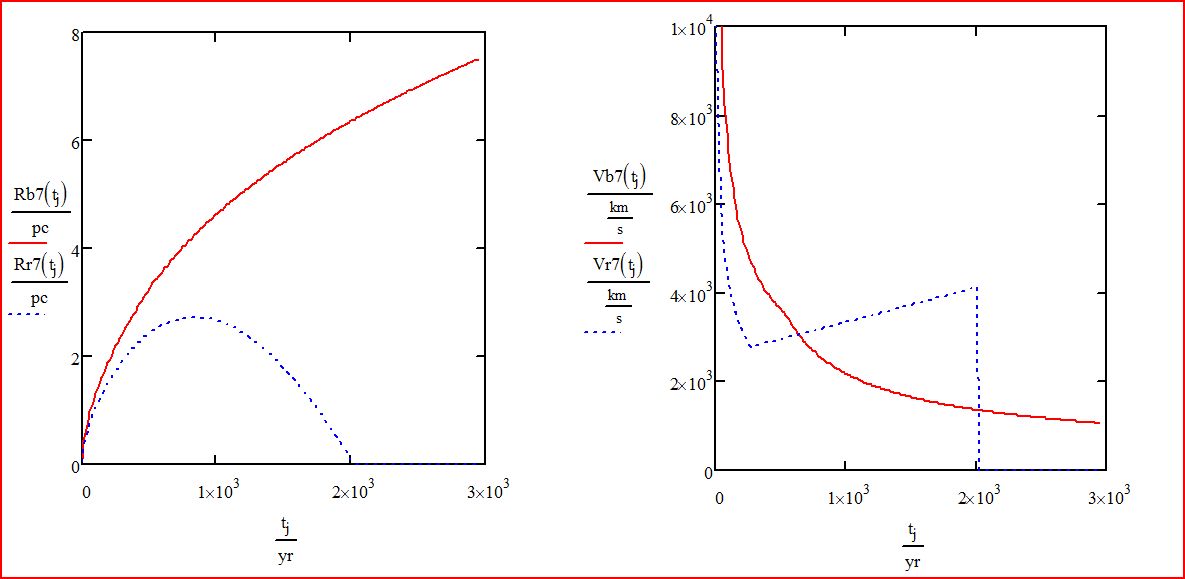} 
%viewport=0 0 636 731
\caption{Left panel: forward and reverse shock radius vs. time for a SNR with
energy $E=10^{51}$erg, ejected mass $2M_{\odot}$, in a uniform circumstellar
medium ($s=0$) with density 1 cm$^{-3}$ and temperature 100 K. 
The ejecta density power-law index is $n=7$. 
Right panel: forward and reverse shock 
velocity vs. time.}
\end{figure}

\begin{figure}
\center
\includegraphics[width=\textwidth]{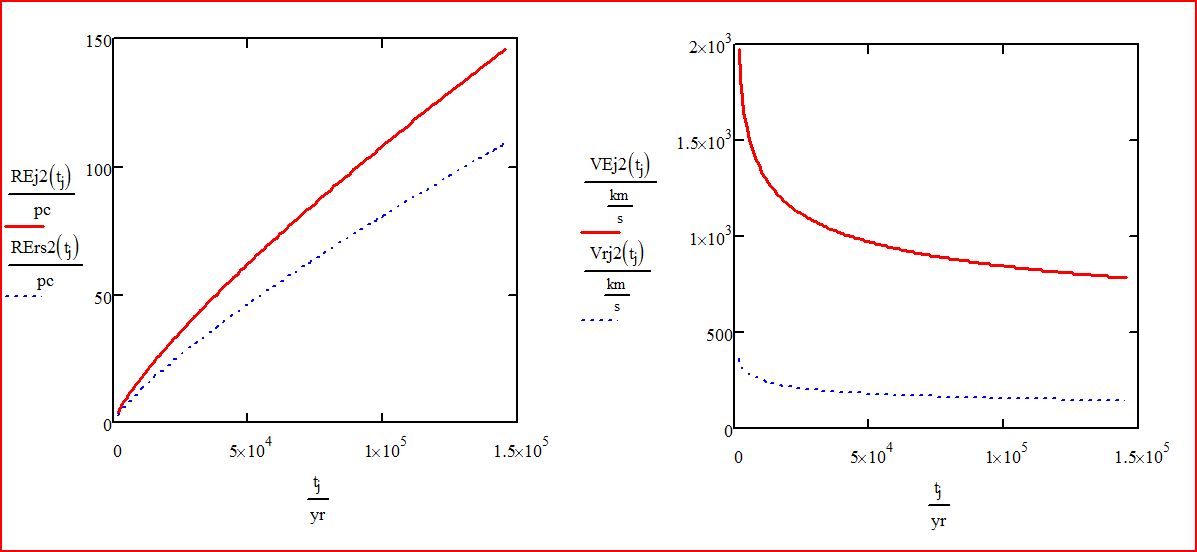} 
%viewport=0 0 636 731
\caption{Left panel: forward and reverse shock radius vs. time for a SNR with
energy $E=10^{51}$erg, ejected mass $2M_{\odot}$, in a stellar wind ($s=2$) with
wind velocity 30 km/s and mass loss rate $10^{-6}M_{\odot}$/yr. The ejecta density power-law index is $n=7$. Right panel: forward and reverse shock 
velocity vs. time.}
\end{figure}

% Do not delete the next line
\small  % Do not delete
%
%%% Comment the following line if you do not have acknowledgments.
\section*{Acknowledgments}   % Do not delete if you declare acknowledgments
%
%%% ACKNOWLEDGMENTS
%%% ACKNOWLEDGMENTS
Support for this work was provided the Natural Sciences and Engineering Research Council of Canada.

%%% REFERENCES
\section*{References}
\bibliographystyle{aj}
\small
\bibliography{proceedings}
Chevalier, R., 1977, ARAA, 15, 175\\
Chevalier, R., 1982, ApJ, 258, 790\\
Truelove, K., \& McKee, C., 1999, ApJSup, 120, 299\\
Taylor, G.I., 1946, Proc.R.Soc.London A, 186, 273\\ 
Sedov, L., 1946, Dokl.Akad.Nauk SSSR, 42, 17\\
Cioffi, D., McKee, C., \& Bertschinger, E., 1988, ApJ, 334, 252\\

\end{document}